# Electron-Phonon Interactions Cause HTSC

J. C. Phillips

Bell Laboratories, Lucent Technologies (Retired), Murray Hill, N. J. 07974-0636, USA

What is the microscopic interaction responsible for high temperature superconductivity (HTSC)? Here data on temporal relaxation of $T_c$ and the room temperature conductivity in $YBa_2Cu_3O_{6+x}$ after abrupt alteration by light pulses or pressure changes are analyzed. The analysis proves, independently of microscopic details, that only electron-phonon interactions can cause HTSC in the cuprates; all other dynamical interactions are excluded by experiment.

## I. INTRODUCTION

After the discovery[1] of HTSC, one of the first questions asked was, are electron-phonon (e-p) interactions responsible for HTSC, just as has been proved[2] for metallic superconductors? Cuprates are poor metals that exhibit, even near optimal doping, residual antiferromagnetism[3], which suggests[4] that magnon exchange could replace phonon exchange as the attractive interaction responsible for Cooper pairing. The only satisfactory theory of microscopically homogeneous superconductivity (the BCS theory) is incompatible with magnetism of any kind[5]. Moreover, when magnetic impurities are added to metallic superconductors, superconductivity is rapidly destroyed.

For metallic superconductors the *isotope effect* has long been considered to provide strong evidence for the critical role played by e-p interactions. However, in HTSC at optimal doping one generally finds only a small isotope effect[6]. This has led to many exotic theories[2] ascribing HTSC to other interactions, especially magnons, or excitons, or even mysterious interactions specified only by some kind of symmetry[3].

There are two pedestrian explanations for the near disappearance of the isotope effect. The first involves the very popular charge-transfer model, which assumes that all that matters is the dopant charge density n which has been transferred to the $CuO_2$ planes from dopants located in other planes. In this simplistic model interactions with the dopants themselves are completely neglected, much as in δ-doped semiconductor interfaces used to study the quantum Hall effect. It is assumed that $T_c = T_c(n)$, and that when $dT_c/dn$ vanishes (optimal doping), the isotope shift $\alpha_Z = -d\ln T_c/d\ln m_Z$ also vanishes. This model has grave weaknesses: in $YBa_2Cu_3O_{6+x}$ after the 60K plateau in $T_c$ for $0.5 < n < 0.7$, where $T_c$ is almost independent of n, $T_c$ rises steeply to ~ 85K for $0.7 < n < 0.8$. This transition is now believed to be associated with ordering of CuO chains in the $CuO_{1-x}$ planes; such ordering really does not change n, except indirectly. Moreover, when this model is applied to both Z = Cu and Z = O isotope data, it leads[6] to numerical discrepancies of a factor of 5 to 8.

The second explanation goes to the opposite limit and emphasizes the dopants, which are supposed to contribute to the formation of conducting nanofilaments, because the material is near a metal-insulator transition analogous to that found in semiconductor impurity bands[7]. There the dopant e-p interactions are very large (about 25x larger than the host e-p interactions), because of the absence of metallic screening in semiconductors. Near the dopants the crystalline symmetry is broken, and neighboring Z ions can relax their positions as part of an anharmonic Jahn-Teller effect. In the cuprates these distortions are especially large, and the free energy is minimized by maximizing the width of a narrow band of high-mobility carriers localized near the dopants and pinned to the Fermi energy. This self-organized, many e-p narrow band explains[8] all the normal-state transport anomalies, which themselves are optimized when $T_c$ is. Isotope replacement slightly alters these complex internal positional adjustments[9], in just such a way as to cancel the changes in the e-p interaction, leaving only a small shift in $T_c$.

This variational Jahn-Teller relaxation model has several attractive features. Because the fraction of Z ions near dopants increases as the dopant density n increases, it explains the general trend of isotope shift decreasing as $T_c$ increases. In LSCO it has recently been discovered[10] that O second neighbors of Sr dopants move *toward* Sr, away from their positions relative to La. As Sr on a La site is effectively negatively charged, and the oxygen ions are also negatively charged, this observation seems to imply an attraction between ions of like charge, contrary to Coulomb's law. However, if these Jahn-Teller displacements are part of the general polarizability relaxation pattern involved in forming the high-mobility band[8], then there is no inconsistency. Note also that when the Jahn-Teller relaxation is so large as to cause substantial energy-level reorderings[10], then it is likely that the resulting strain fields are spread over hundreds of atoms[11].

While the second explanation of the disappearance of the isotope effect in terms of large-scale relaxation of dopant environments may be correct, it is still somewhat unsatisfactory. It seems to depend on many specific quantitative details (normal modes of the large-scale relaxation) to establish a general qualitative conclusion. It raises the question, is there not *some other experiment*, better suited to the complex structures of HTSC, which can test the importance of e-p interactions *directly*? This paper shows that such experiments do exist, and that they lead to simple conclusions regarding the interactions responsible for HTSC which *transcend* all the complex, and not easily measured, details of structure and large-scale relaxation around dopants in these complex materials.

## II. RELAXATION; A SIMPLE AND DIRECT PROBE OF CLASSES OF MICROSCOPIC INTERACTIONS

The alternate experiments involve not merely the details of static local atomic structural relaxation (such as oxygen reordering, or $CuO_6$ octahedral buckling), which is difficult to observe and which cannot be related directly and easily to shifts in $T_c$, but rather *global* dynamical relaxation of the normal-state conductivity and $T_c$ itself. This relaxation has been observed after two kinds of activation, with *light* (photo-induced (super)conductivity)[12] and high *pressures*[13]. The samples were $YBa_2Cu_3O_{6+x}$, with x chosen to maximize either the *light-or pressure*-induced changes. In the *light* case x =

0.4 (near the metal-insulator transition), while in the *pressure* case x = 0.4 - 0.7 (the chain-ordering composition in the CuO$_x$ plane). The data are summarized in Table I.

The observed relaxation of X has been fitted[12,13] by the stretched exponential function

$$\Delta X(t) = \Delta X(0) \exp[-(t/\tau(T))^\beta] \qquad (1)$$

where $\Delta X$ is the shift of the room temperature conductivity $\sigma$ or $T_c$. The relaxation time $\tau(T)$ is thermally activated, with an activation energy $\Delta E_a$ (Arrhenius, 1884). In solids $\Delta E_a$ is generally ~ 1 eV. Note that 1 eV/(valence electron) is a very large energy: it is large enough not only to *melt* the solid, it is also large enough to *vaporize* it. At low temperatures thermal fluctuations of this magnitude are very rare, and that is why, even with an attempt frequency of $10^{12}$/sec, relaxation times at $T_c$ are typically ~ 10 hrs.

At this point the reader is cautioned that traditionally relaxation measurements have enjoyed at best only a mixed reputation. This is both deserved and undeserved. Misgivings are justified because in the older literature (which goes back to Faraday!) many experiments on inhomogeneous samples were fitted indiscriminately with Eqn. (1), and the fitting parameters had no microscopic meaning[14]. However, many modern experiments have been carried out on very carefully prepared samples with highly sophisticated methods, and such experiments should be viewed differently. In the case of the present data, one should note that the internal consistency of the values of $\tau$ and its Arrhenius activation energy $\Delta E_a$ is excellent for the two activation methods, by light and by pressure[15]. Moreover, the fits to (1) span more than four decades in time, leaving little doubt that both $\Delta E_a$ and $\beta$ have been determined[12,13] for YBCO with the same kind of accuracy as has been achieved in several dozens of recent relaxation experiments[14] on a very wide range of homogeneous materials.

### III. TRAPS, DIFFUSION, AND FRACTAL CONFIGURATION SPACE

The key parameter of interest is the dimensionless stretching fraction $\beta = \beta_p$, which turns out to be highly informative. This very surprising and quite unexpected statement must be explained in detail. For a long time it was assumed that Eqn. (1), which was discovered by Kohlrausch about 150 years ago, was only an empirical fitting formula. In the 1970's a number of theorists showed that Eqn. (1) could be derived from a microscopic model. (For references and *many not-obvious fine points* on relaxation, the reader is very well advised to consult the long (75 page) review[14] on this subject.) The model involves diffusion of excitations in a configuration space of dimension $d^*_p$ to randomly distributed traps. As time passes, all the excitations near the traps have disappeared, and only excitations distant from the traps remain. To reach the traps, the latter must diffuse further and further. This leads to (1), which is asymptotic, and also to

$$\beta_p = d^*_p/(d^*_p + 2) \qquad (2)$$

By itself the trap model, with its derivation of (1), is a great accomplishment. Now one should focus attention on (2). At first, it might appear that all that has been done is to replace one empirical parameter, $\beta_p$, with another, $d^*_p$. The systematic review[14], based on ~ 1000 papers (both carefully controlled experiments and powerful numerical molecular dynamics simulations [MDS]), shows that $d^*_p$ is not arbitrary. In fact,

$$d^*_p = d/p \qquad (3)$$

for homogeneous glasses. (The dopants in a well-annealed and homogeneous HTSC presumably form a glassy array, which is homogeneous in YBCO.) Here $d = 3$ is the dimensionality of Euclidean space.

The key factor now is p. Comparison with experiment and several very accurate MDS showed[14] that for homogeneous glasses p is nearly always 1 or 2; it measures the number pd of interaction channels involved in diffusion of excitations in d dimensions. In metals where phonon scattering dominates the resistivity, one of these channels is *always* e-p interactions. However, if other classes of interactions are present, there may be other non-diffusive interaction channels as well. It is easy to see that adding non-diffusive channels increases the stretching factor, which is $1 - \beta_p = 2p/(2p + d)$.

In general, the choice of $p = 1$ or $p = 2$ is dependent on many internal structural factors, and experiment has shown that it is not correct to say (for example) that $p = 1$ trivially for stress and thermal probes, while $p = 2$ for optical and other probes[14]. (For example, in a-Se stress relaxation has been measured macroscopically in a time interval centered on $10^3$ s, and microscopically in time intervals centered on $10^{-9}$ s. The microscopic measurements were carried out by spin-polarized neutron scattering, and as the material is non-magnetic and the a-Se atoms are all equivalent, the microscopic experiment also involves internal stress. The macroscopic *and* microscopic values for $\beta$ are both 3/7, not 3/5, yet in both cases one has stress relaxation, even though the two time intervals differ by a factor of $10^{12}$!) Given the very large value of $\Delta E_a \sim 1$ eV, the reader should not be surprised to hear that, depending on the internal structure, it is easy to activate channels other than the e-p one. Moreover, if these elementary channels contribute to the observable being measured, their activation, and a value of $p > 1$, are almost certain.

### IV. QUASICRYSTALS: A SOLVED, NON-TRIVIAL EXAMPLE

Mathematically the simplest and most rigorous example with $p = 2$ is provided by quasicrystals, where the Euclidean coordinates **r** become $\mathbf{r}_\parallel$, and the Penrose projective coordinates are $\mathbf{r}_\perp$. The Penrose coordinates $\mathbf{r}_\perp$ provide the mathematical constructs which make possible tilings that give rise to five-fold symmetries that are incompatible with normal crystals. Motion in real or $\mathbf{r}_\parallel$ space (the first d channels) involves *phonons* and produces relaxation. On the other hand, motion in $\mathbf{r}_\perp$ space (the second channels) involves *phasons,* which only *rearrange* particles without diffusion or relaxation. In the ideally random quasi-crystal a given hop occurs in the full Penrose configuration space. Thus it may take place along either $\mathbf{r}_\parallel$ or $\mathbf{r}_\perp$. This means that the effective dimensionality $d^*_p$ in which diffusive hopping takes place is given by

$$d^*_p = f_p \, d = d/p \tag{4}$$

where $f_p$ measures the *dimensional effectiveness for actual diffusion* of hopping in pd channels, only d of which are associated with relaxation. For isotropic materials p has nearly always turned out to be either 1 or 2. Although larger values of p are possible in principle, so far they have not been unambiguously observed.

For an axial quasicrystal, which is quasi-periodic only in the plane normal to the axis, the calculation is somewhat more complex. There are five channels, three phonons in $\mathbf{r}_\parallel$ space, and two phasons in $\mathbf{r}_\perp$ space, so that $f_p = 3/5$, and $d^*_p = 9/5$, a rather unusual dimensionality! Thus $\beta_p = d^*_p/(d^*_p + 2) = 9/19$, an even more unusual fraction! However unexpected this fraction may be, it is in *spectacularly excellent agreement* with MDS[16], which gave $\beta_p = 0.47$.

If there are any skeptics left at this point who doubt the validity of the fractal projection procedure, it is suggested that they ask themselves two simple questions: (1) Just how often in the theoretical physics of complex systems has anyone ever achieved agreement like this (no adjustable parameters)? (2) Can such agreement really be coincidental? Are the *dozens of other successes based on extremely high-quality data* that led to this idea[16] also coincidental?

## V. APPLICATION TO HTSC

One can think of phasons as collective coordinates that reflect a long-range interaction that has changed the topological dimensionality of configuration space. Plasmons are another such collective coordinate, this time present in a many-electron system as the result of charge fluctuations and long-range Coulomb interactions. In this case also light-induced charge fluctuations will lead to p ~ 2 (if both phonon and plasmon components of the dopant disorder are glassy). In fact, it is well-known that relaxation of the trap-limited conductivity of amorphous semiconductors can be fitted at low temperatures (glassy limit) by an expression related to (1) that describes dispersive transport[17] with β

~ 0.42(2) ~ $\beta_2$ = 3/7. This value of $\beta$, associated with *two* classes of interactions [charge and density] is very easily distinguished from $\beta_1$ = 3/5 = 0.60 [*only one* class of interaction]. The latter is found, for example, in MDS simulations of relaxation of soft sphere (metallic glass) alloys (no Coulomb interactions)[14,18,19]. The "magic fractions" of $\beta_1$ = 3/5 (phonons only) and $\beta_2$ = 3/7 (phonons *and* plasmons) for the room-temperature photo- and pressure-induced conductivity shifts, shown in Table I, are thus readily explained by the reaction channel theory[14,16] *put forward in 1994*.

Next consider the pressure-induced relaxation of $T_c$ *measured in 1999 (five years later)*. From Table I we see once again the magic fraction $\beta = \beta_1$ = 3/5, indicating that p = 1 and that $T_c$ is shifted *only* by changes in omnipresent e-p interactions. *No other interactions affect $T_c$, as that would lead to p = 2 and $\beta$ = 3/7.* While these 1999 measurements do not refer to x = 0.9, they do show that at x = 0.4, Cooper pairs are formed in *only* the p = 1 (e-p) interaction channel.

The pleasing part of the $\beta_p$ results is their simplicity. It is not necessary to know *any* of the microscopic details of each class of interactions *of any kind* to determine whether or not they are significant. Indeed, for almost all homogeneous glasses, *as stated in 1996 in the abstract to [14],* p = 1 or 2. This includes not only network glasses[20], but also polymers, and many much-studied molecular glass-formers, such as glycerine and ortho-terphenyl[14,21]. The discrete and separable interaction-channel model of Kohlrausch relaxation is essentially universally successful in dozens of these cases, independently of the details of the interactions within a given channel.

The central theme of this paper has been that the interpretation of temporal $T_c$ $\beta$ relaxation data is much simpler than that of isotope shift data $\alpha_Z$ because the latter requires a very detailed microscopic model while the former do not. The situation can be compared to a similar one in the BCS theory[5] for simple metal superconductivity. There the nuclear spin relaxation rate for T just below $T_c$ is enhanced by pair coherence factors, but those factors cancel for the acoustic attenuation rate. Here, if $T_c$ were determined by dynamical pairing through interlayer charge transfer, or by antiferromagnetic exchange interactions, then fluctuations in the corresponding coordinates would raise p from 1 to 2, and reduce $\beta$ from 3/5 to 3/7.

## VI. ISOTOPE SHIFTS ARE COMPLEX IN COMPLEX SOLIDS

Comments on the microscopic interpretation of the isotope shift data $\alpha_Z$ are now appropriate, given the strong evidence that the $\beta$ relaxation data provide for the overwhelming dominance of e-p interactions. Apart from the relaxation effects responsible for cancellation near $T_c$, near the metal-insulator transition (x = 0.4) in YBCO$_{6+x}$, the magnitudes of $\alpha_O$ and $\alpha_{Cu}$ are both close to ½, but $\alpha_{Cu}$ has the *wrong sign* expected from a Debye $\theta$ cutoff in the BCS gap equation for $T_c$. In fact, a good qualitative description of the YBCO data[22,23] is

$$\alpha_O (x) = - \alpha_{Cu} (x) = 0.9 - x \tag{5}$$

In the filamentary model[7], the narrow band of superconductive states pinned to $E_F$ emerges as the result of variational relaxation of atomic and electronic configurations to maximize the polarizability. The filaments are surprisingly well-defined as the paths followed by the centers of masses of hole wave packets in this narrow band, as one can infer from the importance of chain-ordering. One can test the validity of the concept of such literal filamentary paths on an atomic scale in the following way. In YBCO$_{7-\delta}$ near optimal doping the oxygen atoms in the CuO$_{1 \pm \delta}$ plane are ordered into chains, but in cation-substituted alloys L*B*CO = (La$_{1-x}$Ca$_x$)(Ba$_{1.75-x}$La$_{0.25+x}$) Cu$_3$O$_{7+\delta}$ they are not. The degree of of oxygen ordering *changes the metric for percolation in the CuO$_{1 \pm \delta}$ plane in a calculable way.* If one now assumes that the effective length of path segments in this plane must be the same in optimized L*B*CO as in optimized YBCO, then from the optimal value of $\delta = 0.1$ for YBCO$_{7-\delta}$ one can predict the optimal value of $\delta = 0.15$ for L*B*CO$_{7+\delta}$. The sign-reversed result for $\delta$ is in good agreement with experiment[7].

Near the optimized composition, x = 0.9, anharmonic Jahn-Teller relaxation has little effect on the number of coherent filamentary paths. This is because the density of filamentary paths is near its upper limit and it has been optimized. As a result, if an atomic displacement disrupts the phase coherence of a given path, the carriers can adiabatically switch to a new phase-coherent path. However, near the metal-insulator transition, where the paths are few in number and are isolated from each other, the effects of such displacements depend primarily on the degree of internal network organization near weak links. Near the metal-insulator transition at x = 0.4 there are very few chains, and such switching is no longer possible.

One next considers the effects of isotopically modified $\alpha$ atomic displacements for compositions near the metal-insulator transition near x = 0.4. The weak link may be an anion O vacancy. Then large anharmonic Jahn-Teller displacements (similar to zero-point vibrational amplitudes in their mass dependence) of $\alpha$ may take place in order to increase the filamentary conductivity of the high-mobility band by "healing" the weak link. These will draw the $\alpha$ = O electron-rich anions toward the vacancy, but push the $\alpha$ = Cu electron-poor cations away. This can explain the sign reversal in Eqn. (5): the O electron-rich anions increase the density of states at $E_F$ near the weak link much more effectively than the electron-poor cations, and the counter-motion keeps the overall atomic density near the cations nearly constant, thereby reducing internal hydrostatic pressures near the weak link.

## VII. NANODOMAINS ARE PERVASIVE IN PEROVSKITES

Large anharmonic Jahn-Teller effects are observed in many perovskites and pseudoperovskites, including those manganites and nickelates with gigantic magnetoresistance. As shown by experiments on nanocontacts[24], a natural model for these effects involves weak barriers in insulating nanoscale domain walls, which are analogous to weak links in the cuprates, where similar nanoscale antiferromagnetic domain walls are an important part of the filamentary model[7]. In such cases it is possible to have large isotope shifts of either sign, even with O isotopes alone. This has been observed[25] for the metal-insulator transition temperature, with negative $\alpha_O$ between 0 and -0.9 for $RNiO_3$ for a number of rare earths R, and with positive $\alpha_O$ between 0 and 0.9 for $La_{1-x}Ca_xMnO_3$ with x in the very narrow interval between 0.2 and 0.3. Recent experiments[26] have also shown persistent photoconductivity in manganites similar to that seen in the cuprates; these can also be explained by large Jahn-Teller effects.

As recently stressed by Muller, Bednorz and he originally studied[1] doped $La_2CuO_4$ hoping for higher $T_c$'s through large electron-phonon interactions associated with large Jahn-Teller distortions, whose anharmonicity may be indicated by large isotope shifts. Such shifts continue to be observed in many properties[27,28]. However, the presence of such shifts does not completely exclude the active participation of magnons as well as phonons in forming Cooper pairs.

## VII.  CONCLUSIONS

Here it has been shown that there is room in the $T_c$ $\beta$ relaxation data, measured independently by two different probes, only for e-p interactions in the *electrically active* (probably filamentary) regions of the superconductive phase. Local structural relaxation can conceal, and even cancel, $\alpha$ isotope effects of a microscopic dynamical nature at optimized doping. Local dynamical correlations associated with microscopic filamentary formation can reverse the sign of $\alpha$ isotope effects between cations and anions. By contrast, the macroscopic temporal relaxation $\beta$ data are not subject to such cancellations, and in fact the agreement between theory and experiment for $\beta$ is much better than that for $\alpha$, even in the old (low $T_c$) superconductors. The data show that *electrically active* carriers become superconductive at $T_c$ in YBCO *only because of e-p interactions*, which is consistent with fundamental microscopic theory[5]. The simplicity of the interpretation of the global $\beta$ relaxation data also helps one to construct microscopic models for the very simple, but hitherto quite puzzling, trends and sign reversals in the $\alpha_Z$ isotope shifts, not only in cuprates, but also in gigantic magnetoresistance perovskites.

I am grateful to Prof. K. A. Muller for encouraging discussions.

# APPENDIX: THE GAPS BETWEEN MATERIALS SCIENCE, STATISTICAL PHYSICS, AND MICROSCOPIC THEORIES OF HTSC

Many readers have experienced difficulty in understanding this paper, some because they disagree with the conclusions on political grounds. Others have struggled because the mathematical theory of stretched exponential relaxation discussed here was developed in the 1970's, long before there were relaxation experiments on good samples, or sufficiently large-scale numerical simulations. As a result, the general theory of relaxation, and its implications for the nature of microscopic interactions, has been largely ignored, and to a great extent even forgotten – an object lesson to theorists concerning the importance of what patent lawyers call "reduction to practice". There is also a very large interdisciplinary gap between highly specialized experimentalists, who are sufficiently skilled in the intricacies of materials science to produce high quality data on HTSC, on the one hand, and equally highly specialized theorists, who construct very abstract models of complex systems, such as quasicrystals. Finally, there is a third group of theorists who are attempting to construct microscopic theories of HTSC without knowing much about *either* materials science *or* statistical physics, and so are quite unprepared to deal with the glassy statistical complexities that are an unavoidable part of the physics of dopant-disordered HTSC.

. It could be argued that in principle the general reader would not encounter these problems if he had the time needed to read the very lengthy review[14] on relaxation, but in reality, few have the luxury to do so. Indeed, the history of the theory and practice of relaxation science is a fascinating topic in its own right. The subject traces its roots to the properties of ionic relaxation in electrolytes, as studied by Faraday and Kohlrausch ~ 1850, and partially explained by Arrhenius ~ 1880, in classic work that earned him the Nobel prize in 1903, and established him as the father of all theories of wet chemistry.

The crossover from this "prehistoric" era to modern theory occurred with S-like progress, centered on the 1980's. This progress was stimulated by theoretical work[14] by physicists and mathematicians in the 1970's. However, that beautiful and elegant work has never been accepted by the polymer and molecular chemists, who are responsible for most of the experimental data. The reason for this is that the measured values of $\beta$ sometimes reflect simply relaxation of density fluctuations, in cases where the influence of internal structure is small (p = 1, $\beta$ = 3/5). However, equally often they reflect relaxation in hyperspace (when there is strong internal structure, as in polymers, or long-range Coulomb forces), which gives p = 2, $\beta$ = 3/7. Readers who still are interested in combining theory and experiment to advance their understanding of HTSC, in order eventually to reach a microscopic theory analogous to the BCS theory of older superconductors, may not have time to follow all these twists and turns (however entertaining). They should consider the following points.

(1) It is highly desirable to have an alternative and very simple and direct approach to the isotope effect for probing the nature of the microscopic interactions in perovskites and related oxides, just because these materials are so complex. These interactions have given rise to such striking phenomena as not only HTSC, but also the closely related phenomenon of colossal magnetoresistance (CMR). There have been hundreds of experiments on the isotope effect in HTSC, but (as discussed in Sec. 6) the results, although large and even spectacular in many cases, have been generally described as mysterious. The price we have to pay for replacing the mysterious isotope effect parameter $\alpha$ by the explicable relaxation parameter $\beta$ is small, in the sense that $\beta$ can be understood in terms of the universal theory that has already been worked out and confirmed in detail for other materials. On the other hand, with large unit cells and many internal degrees of freedom, it appears to be essentially impossible to develop a similar theory for the isotope effect parameter $\alpha$. The qualitative results for $\alpha$, such as the disappearance of the effect on $T_c$ near optimal doping, and the sign reversal for Cu compared to O, are explicable in the topological filamentary model, but this itself, although intuitive, is mathematically even more sophisticated than relaxation theory. In other words, there is no cheap and easy way to understand HTSC, and if one takes the subject seriously, one must be prepared to "pay one's dues" to understand it.

(2) The relaxation parameter $\beta$ provides just such an alternative and very *simple and direct* approach for probing the nature of the microscopic interactions. Its utility has been demonstrated in dozens of cases, and it is easy to see, from studying the historical evolution of the data base[14], that rapid advances in experimental techniques have greatly improved the agreement between theory and experiment. For homogeneous samples the theory works equally well in all cases, *including HTSC*. There is no need to decompose these interactions into high frequency or low frequency ones, interlayer or intralayer ones, etc., as they are naturally sorted into separate and countable channels by the nature of configuration space. Some have argued that relaxation takes place slowly, and therefore trivially involves only phonons. One person even suggested that "The fact that the relaxation of low-energy excitations is caused by phonons is a trivial consequence of the conservation of energy." This is nonsense. At the molecular level the activation energy for relaxation $\Delta E_a$ typically is of order 1 eV in all cases studied, *including HTSC*. This activation energy is by no means small: in fact, it is large enough to include not only phonons, but also plasmons, magnons, and any other kind of past or present, real or fictional, whaton: in other words, *all* the microscopic interactions (density, charge and spin fluctuations) that might conceivably contribute to formation of Cooper pairs, might also contribute to relaxation of $T_c$. This is the point that should be obvious, and truly trivial, to both materials scientists and to mathematical and statistical theorists. The way the two quantities, HTSC and relaxation, are affected by microscopic interactions in qualitatively different channels is, of course, quite different. For coherent Cooper pairing, the competition between phonons and magnons is destructive and difficult to estimate for complex materials, while for incoherent relaxation the situation is remarkably simple. The pure number $\beta$ is a *direct measure of the total number (no cancellations!) of such interactions*, which is independent of all microscopic details, such as static or dynamical coupling constants (providing only that the defects in the samples are strongly disordered).

(3) Another favorite objection concerns the agreement between theory and experiment. This is actually *much, much better (almost perfect!)* for the relaxation parameter $\beta$ in complex HTSC than it has ever been for the isotope effect parameter $\alpha$, even in the classically simple case of elemental metals (at best agreement within 20%). Because of numerous cancellations the difficulties in analyzing the effects of isotopic substitution on $T_c$ are impossibly large for HTSC.  Isotope effects on other quantities, such as pseudogaps, have been obtained by various kinds of magnetic resonance, but the analysis required for these is far greater still, and although such data often exhibit very large effects indicative of giant e-p interactions, their complete meaning may never be successfully and quantitatively interpreted at the microscopic level.  In the absence of such interpretations, which have never even been suggested by anyone, to say, as some have done, that such experiments could falsify the model discussed here, is ludicrous.

(4) Sometimes it is suggested that such a (relatively!) simple theory as the present one is inadequate because it does not explain everything.  In general such arguments smack of sophistry (one of the favorite debating techniques of sophists, when they were losing an argument, was to change the subject).  For HTSC the situation is even worse: there are more than 50,000 experimental papers, and 20,000 theoretical ones, and only a handful of the latter attempt to discuss experiment at all.  In a very large fraction (~ 95% or more) of the experiments, the discussion is not accompanied by any serious, perovskite-specfic, BCS-compatible microscopic theory.  In these circumstances, a simple, direct, and reliable alternative to the isotope effect, which does not suffer from all the weaknesses described above, should be much sought-after, and highly prized.

| Pump | Probe | x | T(K) | $\beta_{Exp}$ | $\beta_{Th}$ | Ref. |
|---|---|---|---|---|---|---|
| 2 eV photon [1] | σ | 0.4 | 293 | 0.43 | 3/7 | 12 |
| 2 eV photon [2] | σ | 0.4 | 293 | 0.58(2) | 3/5 | 12 |
| Δp = 0.7GPa | σ | 0.4, 0.7 | 298 | 0.60(1) | 3/5 | 13 |
| -Δp = (0.3 – 0.8)GPa | $T_c$ | 0.4 | 40 | 0.60(8) | 3/5 | 13 |

Table I. Relaxation parameters measured with light and pressure pumps. [1] Relaxation of saturated photoconductivity after prolonged illumination, measured at annealing or resting (see ref. 14, p. 1181) room temperature. [2] Build-up of photoconductivity during prolonged illumination at constant intensity. The quoted value of $\beta_{Exp}$ weights the samples with $\Delta\sigma/\sigma_0 < 1/3$ twice as heavily as those with $\Delta\sigma/\sigma_0 > ½$ in order to avoid non-linear effects.

Email: jcphillips@lucent .com

1. J. G. Bednorz and K. A. Muller, Z. Phys. B **64**, 189 (1986).
2. J. C. Phillips, *Physics of High-$T_c$ Superconductors* (Academic Press, Boston,1989).
3. D. Vaknin, J. L. Zarestky, and L. L. Miller, Physica C **329**, 109 (2000).
4. V. J. Emery, Phys. Rev. Lett. **58**, 2794 (1987).
5. J. Bardeen, L. N. Cooper, and J. R. Schrieffer, *Phys. Rev.,* **108**, 1175 (1957); I. E. Dzyaloshinskii, and A. I. Larkin, *Sov. Phys. JETP,* **34**, 422 (1972).
6. J. P. Franck, Physica C **282-287**, 198 (1997); Phys. Scrip. T**66**, 220 (1996).
7. J. C. Phillips, Sol. State Comm. **109**, 301(1999); Phil. Mag. B **79** 527, 1477 (1999).
8. H. L. Stormer, A. F. J. Levi, K. W. Baldwin, M. Anzlowar, and G. S. Boebinger, Phys. Rev. B Solid State **38**, 2472 (1988).
9. D. Louca, G. H. Kwei, B. Dabrowski, and Z. Bukowski, Phys. Rev. B **60**, 7558 (1999).

10. D. Haskel, E. A. Stern, D. G. Hinks, A. W. Mitchell, and J. D. Jorgensen, Phys. Rev. B **56**, R521 (1997).

11. S. Ogut and J. R. Chelikowsky, Phys. Rev. Lett. **83**, 3852 (1999).

12. V. I. Kudinov, I. L. Chaplygin, A. I. Kirilyuk, N. M. Kreines, R. Lahio, E. Lahderanta, and C. Ayache, Phys. Rev. B **47**, 9017 (1993).

13. S. Sadewasser, J. S. Schilling, A. P. Paulikas and B. W. Veal, Phys. Rev. B **61**, 741 (2000).

14. J. C. Phillips, Rep. Prog. Phys. **59**, 1133 (1996); J. C. Phillips and J. M. Vandenberg, J. Phys.: Condens. Matter **9**, L251-L258 (1997).

15. J. C. Phillips, Physica C **340**, 292 (2000).

16. J. C. Phillips, J. Non-Cryst. Sol. **172-174**, 98(1994); M. Dzugutov and J. C. Phillips, J. Non-Cryst. Solids **192-193**, 397 (1995).

17. C. E. Nebel and G. H. Bauer, Phil. Mag. B **59**, 463 (1989).

18. J. N. Roux, J.- L. Barrat, and J.-P. Hansen, J. Phys.: Cond. Matt. **1**, 7171 (1989).

19. H. Teichler, Phys. Stat. Solidi b **172**, 325 (1992).

20. J. C. Phillips, *Rigidity Theory and Applications* (Ed. M. F. Thorpe and P. Duxbury, Mich. State Univ. Press 1999), p.155; P. Boolchand, X. Feng, D. Selvanathan, and W. J. Bresser, *ibid.,* p.279.

21. R. Bohmer, G. Hinze, G. Diezemann, B. Geil, and H. Sillescu, Europhys. Lett. **36**, 55 (1996).

22. D. E. Morris, A. P. B. Sinha, V. Kirtikar, and A. N. Inyushkin, Phys. C **298**, 203 (1998).

23. J. P. Franck, J. Jung, M. A.-K. Mohamed, S. Gygax and G. I. Sproule, Phys. Rev. B **44**, 5318 (1991).

24. G. Tatara, Y.-W. Zhao, M. Munoz, and N. Garcia, Phys. Rev. Lett. **83**, 2030 (1999).

25. M. Medarde, P. Lacorre, K. Conder, F. Fauth and A. Furrer, , Phys. Rev. Lett. **80**, 2397 (1998).

26. A. Gilabert, Proc. SPIE Conf. 4058 (to be published).

27. K. A. Muller, Physica C **341-348**, 905 (2000).

28. D. R. Temprano, J. Massot, S. Janssen, K. Conder, A. Furrer, H. Mutka and K. A. Muller, Phys. Rev. Lett. **84**, 1990 (2000).